\documentclass[preprint,12pt]{elsarticle}

\usepackage{amssymb}
\usepackage{graphicx}
\usepackage{textcomp} 

\biboptions{sort,compress}

\journal{Int. J. Mass Spec.}

\begin{document}

\begin{frontmatter}



\title{A large Bradbury Nielsen ion gate with flexible wire spacing based on photo-etched stainless steel grids and its characterization applying symmetric and asymmetric potentials}


\author[triumf,tum]{T.~Brunner}
\ead{thomas.brunner@triumf.ca}
\author[stanford]{A.R.~Mueller}
\author[stanford]{K.~O'Sullivan}
\author[triumf]{M.C.~Simon}
\author[triumf]{M.~Kossick}
\author[triumf,ubc]{S.~Ettenauer}
\author[triumf,ubc]{A.T.~Gallant}
\author[triumf]{E. Man\'e}
\author[triumf]{D. Bishop}
\author[triumf]{M. Good}
\author[stanford]{G.~Gratta}
\author[triumf,ubc]{J.~Dilling}
\address[triumf]{TRIUMF, 4004 Wesbrook Mall, Vancouver, V6T 2A3, Canada}
\address[tum]{Physik Department E12, Technische Universit\"at M\"unchen, James Franck Stra\ss e, D-85748 Garching, Germany}
\address[stanford]{Physics Department, Stanford University, Stanford, CA, USA}
\address[ubc]{Department of Physics and Astronomy, University of British Columbia, 6224~Agriculture Road, Vancouver, BC, V6T 1Z1, Canada}

\begin{abstract}
Bradbury Nielsen gates are well known devices used to switch ion beams and are typically applied in mass or mobility spectrometers for separating beam constituents by their different flight or drift times. A Bradbury Nielsen gate consists of two interleaved sets of electrodes. If two voltages of the same amplitude but opposite polarity are applied the gate is closed, and for identical (zero) potential the gate is open. Whereas former realizations of the device employ actual wires resulting in difficulties with winding, fixing and tensioning them, our approach is to use two grids photo-etched from a metallic foil. This design allows for simplified construction of gates covering large beam sizes up to at least 900\,mm$^2$ with variable wire spacing down to 250\,\textmu m. By changing the grids the wire spacing can be varied easily. A gate of this design was installed and systematically tested at TRIUMF's ion trap facility, TITAN, for use with radioactive beams to separate ions with different mass-to-charge ratios by their time-of-flight.
\end{abstract}


\end{frontmatter}


\section{Introduction}
\label{intrto}
\begin{figure}
   \begin{center}
   \includegraphics[width=1\textwidth]{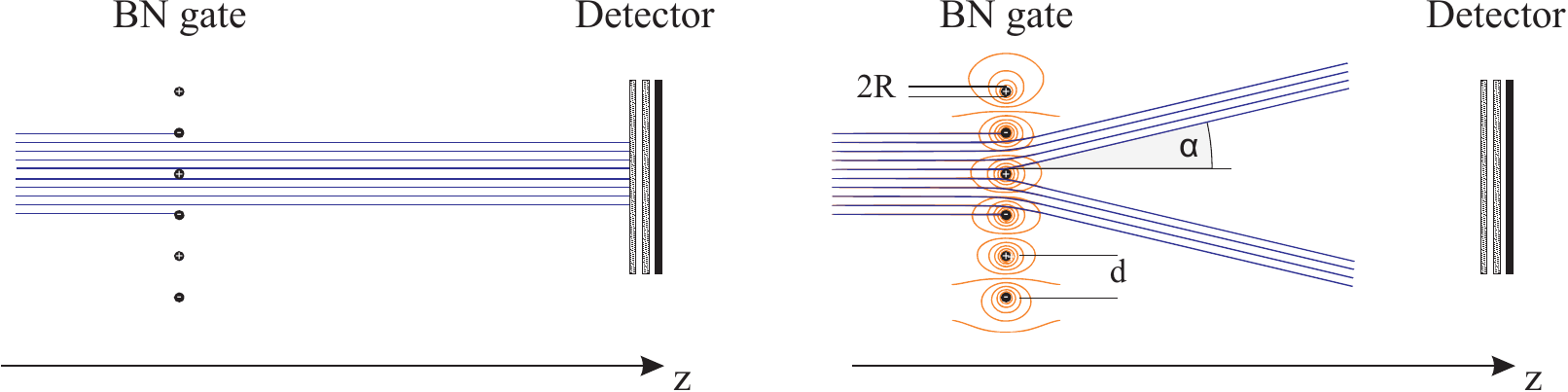}
   \end{center}
  \caption{Schematic of the working principle of a BN gate showing idealized ion trajectories. If both wire sets are at the same potential, most of the ions cross the gate and reach the detector (left). If opposite voltages are applied to the wires, the beam is deflected, effectively turning off the gate (right). Also shown are equi-potential lines of the wires.}
 \label{fig:principle}
\end{figure}

In many applications such as ion mass or ion mobility spectrometers, it is necessary to create pulses from continuous ion sources~\cite{Laa10}. Similar requirements apply to particle accelerators, where beam pulses need to have specific temporal profiles. Moreover, at radioactive isotope facilities such as CERN-ISOLDE~\cite{Kug00} and TRIUMF-ISAC~\cite{Dom00} one typically suffers from isobaric contamination that can be suppressed by isolating the desired rare isotope by its time of flight~\cite{Rod10b,Wol11}.
The use of Bradbury Nielsen (BN) gates~\cite{Bra39} is a common way to realize short ion pulses. These devices require lower voltages, can be more compact and exhibit lower capacitance than alternatives such as deflection plates. The original design of a BN gate consists of two sets of parallel, interleaved wires, with the wire plane perpendicular to the ion beam axis~\cite{Bra39}. In the case of identical potential on both wires, generally chosen to coincide with the beam line ground potential, the gate is open and most of the ions penetrate undisturbed. Once a voltage of opposite polarity but equal amplitude is applied, the ions are deflected by an angle~$\alpha$ as defined in Fig.~\ref{fig:principle}. The deflection angle $\alpha$ for round wires is given by\footnote{Strictly speaking the formula is only correct for deflection voltages applied during the deflection of the ion. The field geometry from a BN gate makes corrections to this deflection angle quite small.} \cite{Yoo05}:
\begin{equation}
 \tan\alpha=\frac{\pi}{2\,\ln\left(\cot\left(\frac{\pi\,R}{2\,d}\right)\right)}\,\frac{V_{wire}}{E_{kin}/q}.
\label{eq:alpha}
\end{equation}
The deflection angle depends on the wire diameter $2\cdot R$, the distance between wires $d$ and the ratio of voltage applied to the wires $V_{wire}$ ($V_{pos}=\left|V_{neg}\right|$) over the ion's kinetic energy $E_{kin}$ and the ion charge state, $q$. $\alpha$ decreases when the ion is in the vicinity of the gate's electric field while the potential is switched.
Depending on the requirements of the system these parameters can be set so that the ions are deflected in such a way that they cannot reach the detector (a state referred to as \textit{gate closed} in this paper). Deflected ions are deposited somewhere along the beam line. For a detailed discussion on the deflection angle $\alpha$ see \cite{Yoo05}.

An important property of BN gates is that they only influence the motion of the beam at distances comparable to the wire spacing $d$. This results in a compact design and fast switching, providing a close to ideal definition of a \textit{gate} in contrast to deflection (kicker) plates.
The geometry of BN gates generally also results in smaller capacitance, simplifying the drive circuitry and further aiding the goal of fast switching.
More importantly, the spatial dimensions of the gate along the beam axis and thus the disturbance of the ion's flight path is greatly reduced for BN gates compared to kicker plates. 

BN gates of large (cm) size are typically built by stretching wires on frames in various arrangements~\cite{Vla96,Kim01,Szu05,Kyu}. This results in substantial complexity and delicate devices that are difficult to assemble and not reliable. This is particularly true when only clean and ultra-high vacuum components can be used. In order to overcome these drawbacks, a new type of BN gate design has been developed based on chemically etched wires. The new design simplifies the construction while at the same time providing far more robust and reliable assemblies. Additionally, the size of the gate is easily scalable, covering large active areas. Currently, an active area of 900\,mm$^{2}$ has been realized where the wire spacing can easily be changed simply by replacing the grids.  
\section{Gate Design}
\begin{figure}
   \begin{center}
\includegraphics[width=.6\textwidth]{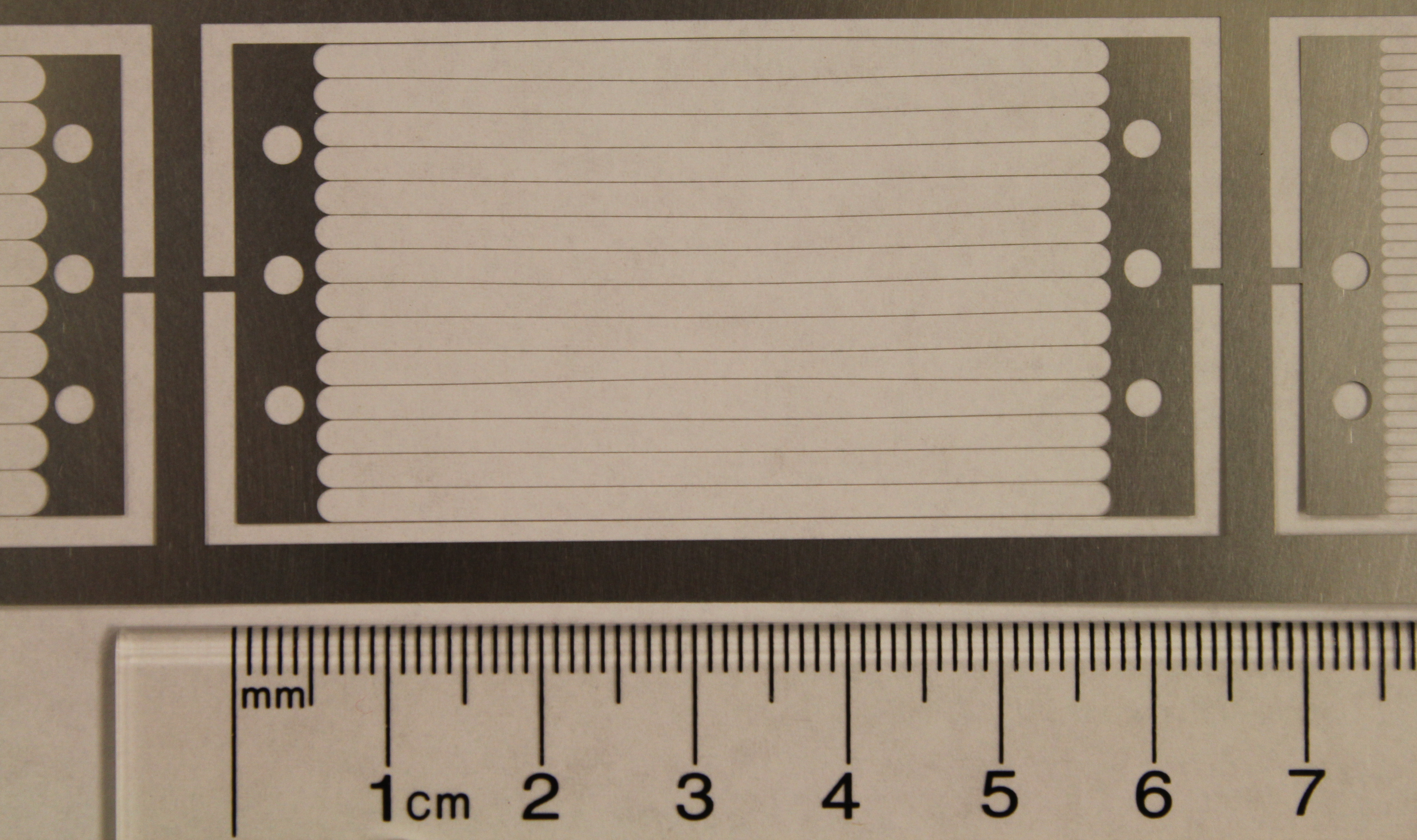}
   \end{center}
  \caption{A photo-etched grid for a BN gate. Note that multiple wire spacings (three in this case) can be obtained on the same etched foil. }
 \label{grid}
\end{figure}

The critical part in assembling a BN gate is the precise positioning of the two sets of wires isolated from each other. In the design described here, these issues are overcome by using photo-etched grids, as shown in Fig.~\ref{grid}. The basic idea is to handle two wire grids instead of individual wires. 
In addition, uniform wire tensioning is automatically achieved and the handling of the grids is substantially easier than that of individual wires.

The wire grids described here were photo-chemically etched from a 50.8\,\textmu m thick stainless steel sheet by Newcut Inc.~\cite{newcut}. Thus, the resulting wires exhibit the approximate cross section of a diamond with similar dimensions of 50\,\textmu m in both thickness and width\footnote{The tolerances of metal sheet thickness and wire diameter are 10\,\% and 13\,\textmu m, respectively~\cite{Engel}. The wire diameter cannot be smaller than the thickness of the metal sheet.}. While this leads to a field configuration that is different from that of ideal, circular wires, such an effect is only important near the surfaces and we do not expect it to alter the general behavior of the gate, described by Eq.(\ref{eq:alpha}). This is because of the large spacing as compared to the wire diameter. Mounting holes were located on each side of the wire grids. The grid-to-grid alignment is provided by the position of the mounting holes. In our design the grid is centered by the mounting screws. If more precise tolerances are required, the positioning of the wire grids could be realized by alignment pins. In order to be able to use identical mounting frames for the two grids, the mounting holes are offset with respect to the symmetry axis of the grid. 

\begin{figure}
   \begin{center}
   \includegraphics[width=.9\textwidth]{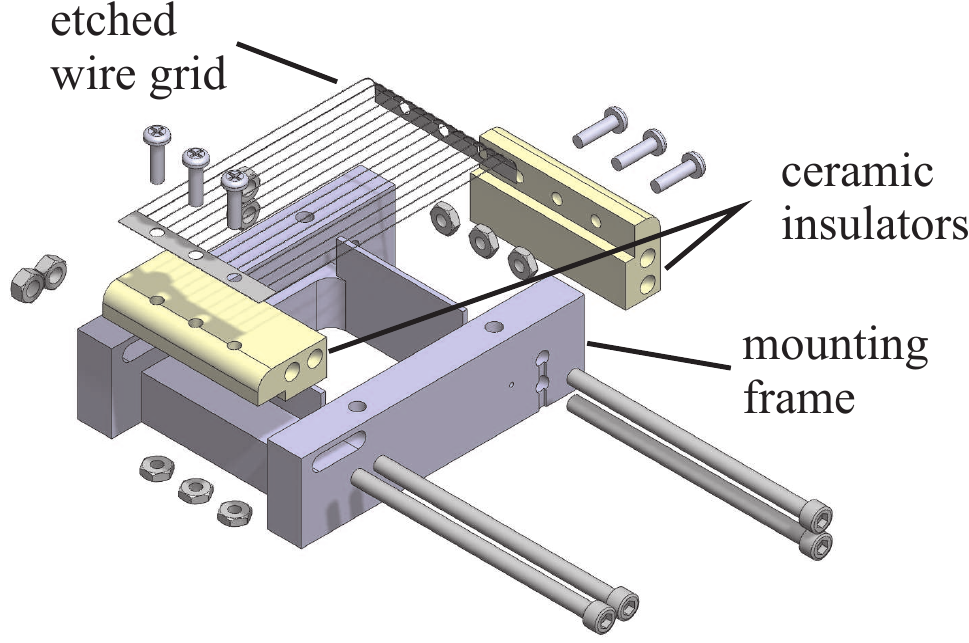}
\includegraphics[width=.9\textwidth]{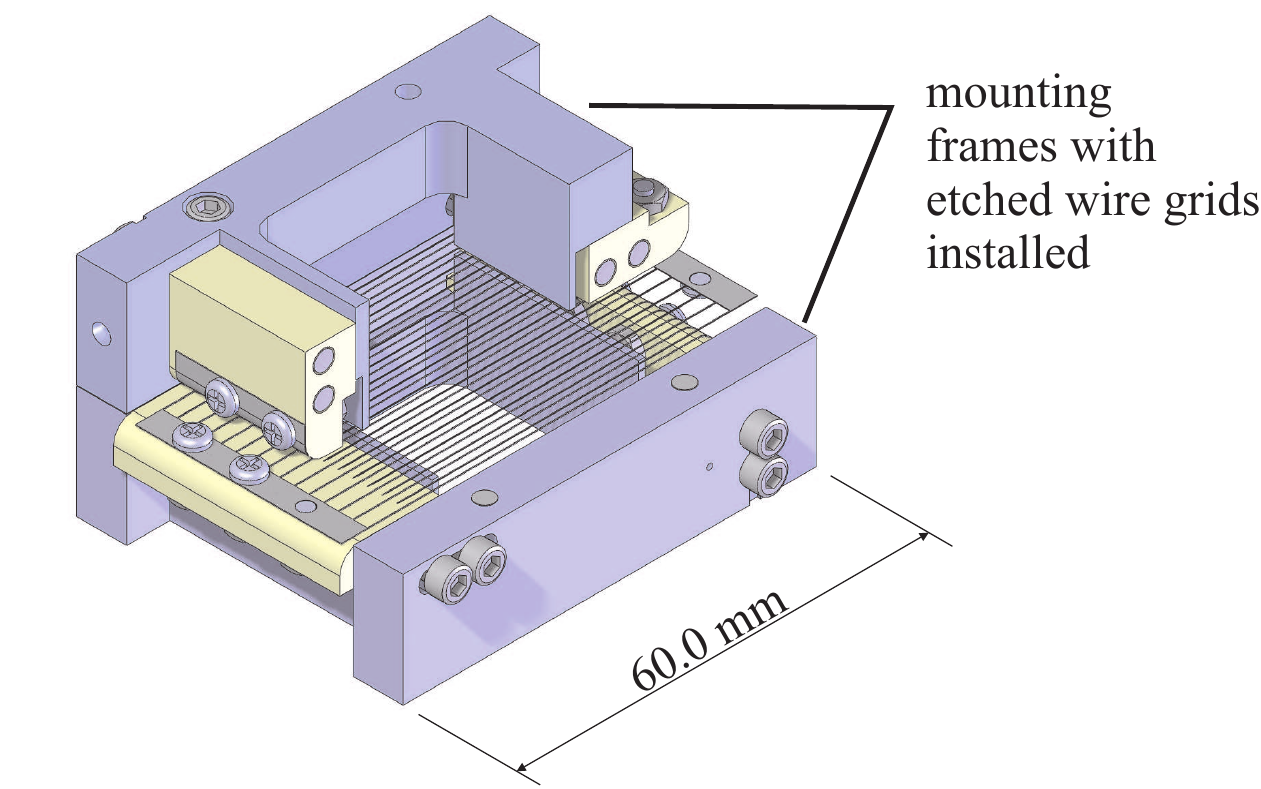}
   \end{center}
  \caption{Exploded view of one half BN gate showing stainless steel mounting frame structure, wire grid and mounting ceramics (top). A section view of the fully assembled BN gate consisting of two identical frame structures rotated by $180^{\circ}$ to each other corresponding to the two polarities (bottom).}
 \label{fig:bngass}
\end{figure}
\begin{figure}
   \begin{center}
   \includegraphics[scale=0.5]{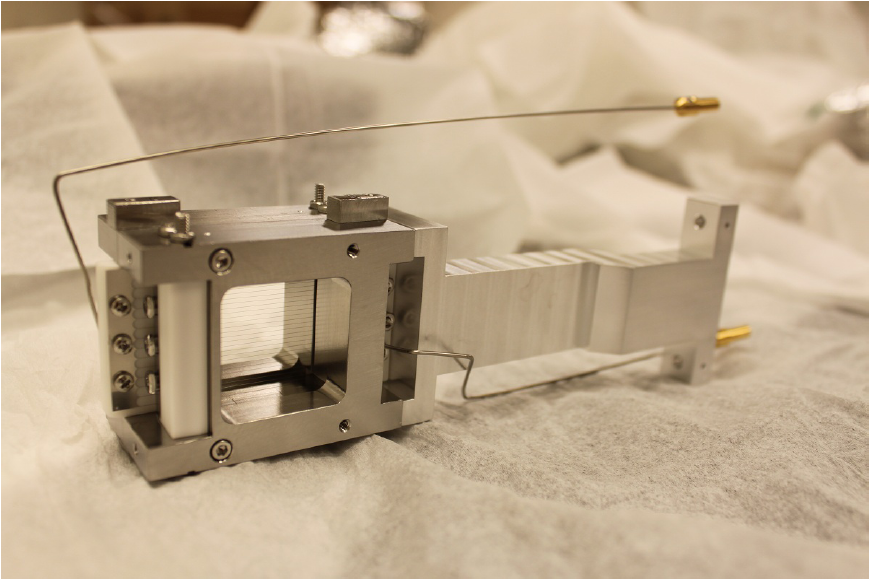}
   \end{center}
  \caption{A fully assembled BN gate with mounting bracket and electrical connections. This particular gate had $d=1.1$\,mm wire spacing and a total aperture of 900\,cm$^2$.}
 \label{fig:realgate}
\end{figure}

Each grid is mounted onto a pair of macor isolators that is in turn installed onto a stainless steel frame, as shown in Fig.~\ref{fig:bngass}. The tension of the grid is set by the screws holding the macor blocks onto the frame and is easily adjusted at the time of assembly. A dedicated tensioning screw pushing the macor blocks apart was found unnecessary for the sizes tested. A rounded edge of the macor blocks allows each grid to have one end bent out of the way, so that when the two frames are mounted against each other, the wires are interleaved and the two grids are not shorted with each other. For simplicity of fabrication, grids, macor blocks and frames are all identical and assembled in a mirrored configuration. A picture of a fully assembled and mounted gate is presented in Fig.~\ref{fig:realgate}. 

The large extension of the gate body in beam direction protects the grids and acts as the beam dump if the gate is closed. While stainless steel was used to fabricate the grids described here, other metals can be used as well. Because of the intrinsically clean construction materials (macor and stainless steel) a vacuum of $2\times 10^{-10}$\,mbar was achieved during all measurements.
\section{Characterization}
\begin{figure}
   \begin{center}
   \includegraphics[width=.8\textwidth]{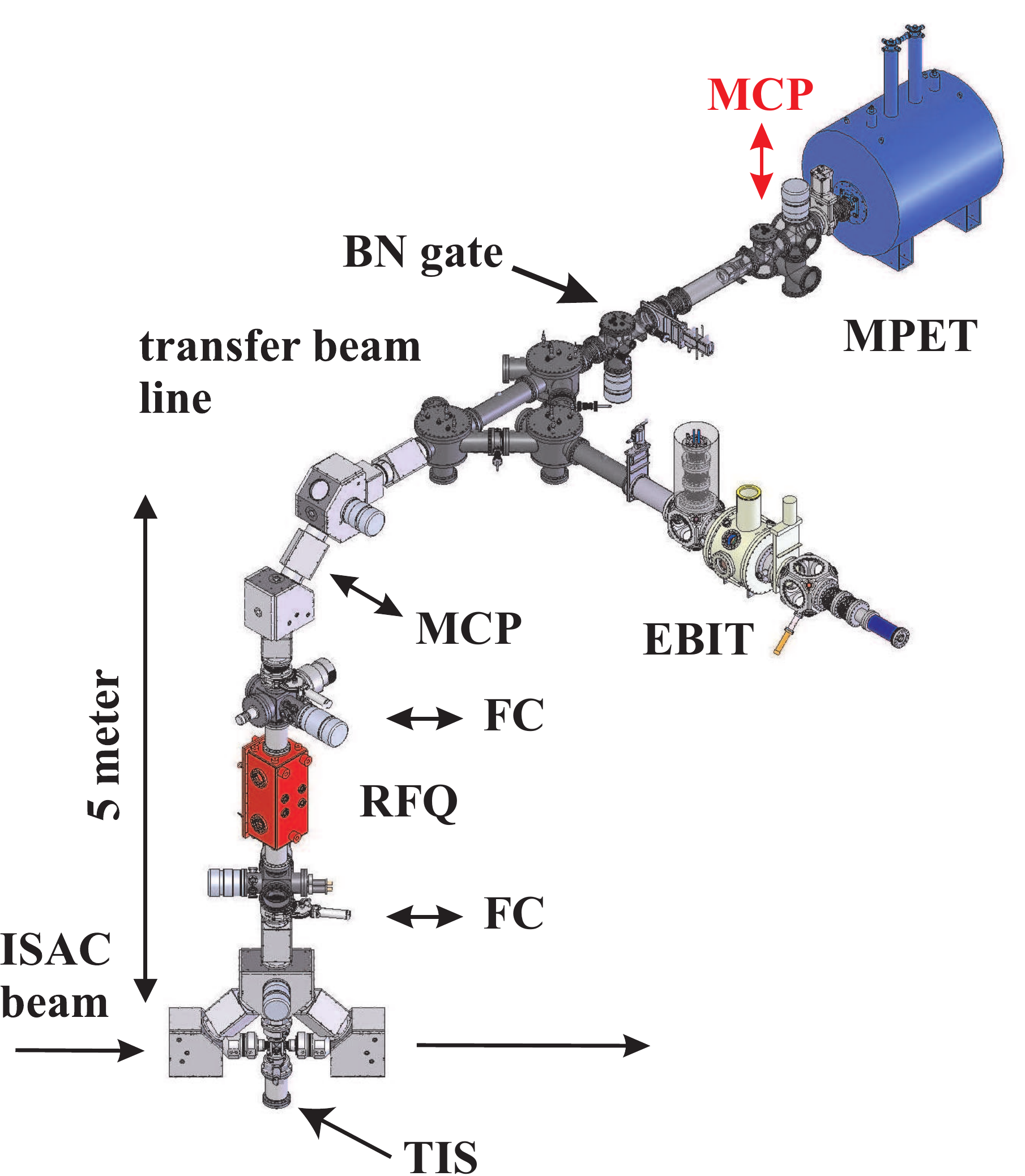}
   \end{center}
  \caption{Schematic of the TITAN setup. For the characterization of the BN gate, the RFQ was delivering a continuous beam of K$^+$/Rb$^+$ ions at an energy of 3\,keV. The ion beam was delivered by the test ion source (TIS)~\cite{Bru11a}. MCP detectors and Faraday cups (FC) can be moved into the beam line while the BN gate is installed permanently. The position of the MCP detector used in the presented studies is indicated by a red arrow.}
 \label{fig:BNG-TITAN}
\end{figure}
\begin{figure}
   \begin{center}
   \includegraphics[width=1\textwidth]{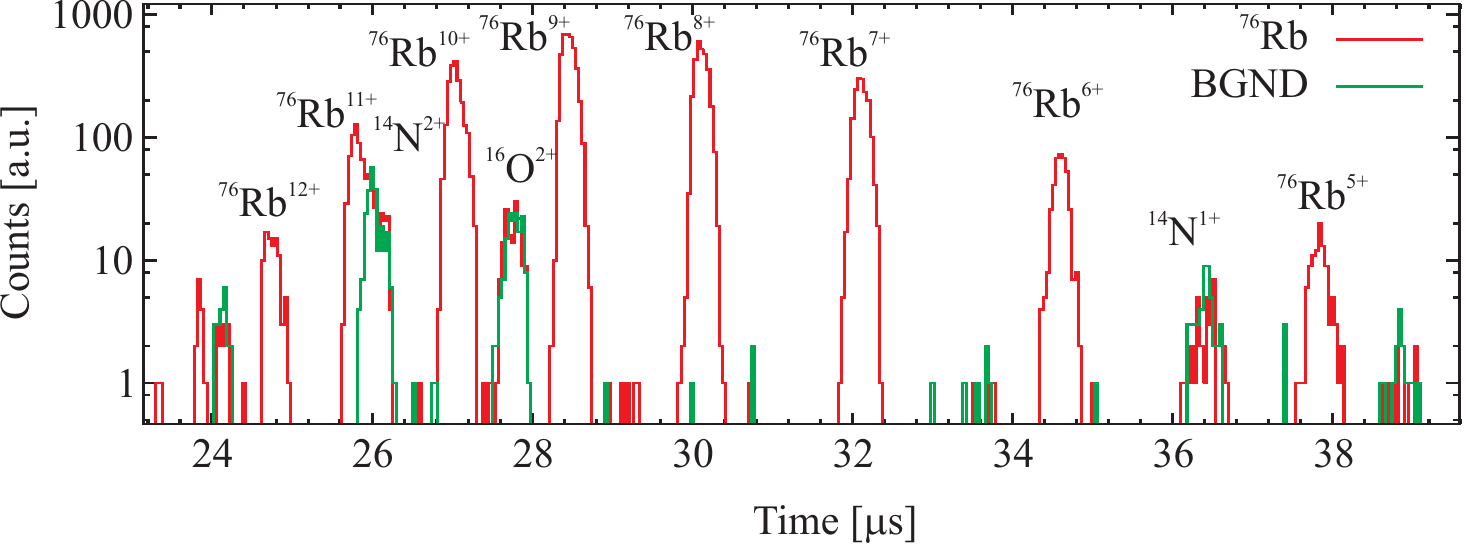}
   \end{center}
  \caption{Time-of-flight distribution of charge-bred radioactive $^{76}$Rb (red spectrum). The ion bunch was extracted from the EBIT and detected with the same MCP detector used for the characterization presented in this work. The background contamination originating from the EBIT is also displayed (green spectrum).}
 \label{fig:ToFRb76}
\end{figure}
The presented large-area BN gate was installed at TRIUMF's Ion Trap setup for Atomic and Nuclear science (TITAN)~\cite{Dilling2006198}. The TITAN setup consists of several ion traps and is dedicated to high precision measurements of radioactive, short lived nuclei (see e.g. \cite{Smi08,Ryj08,Rin09,Lap10}). A schematic of the TITAN facility is presented in Fig.~\ref{fig:BNG-TITAN} indicating the location of the BN gate. Radioactive isotopes are produced by bombarding an ISOL type~\cite{Aey08} target at the Isotope Separation and ACceleration (ISAC) facility \cite{Dom00} with 500\,MeV protons. During typical operation, the BN gate is used to separate different isotopes by their time-of-flight when ejected from the first ion trap within the TITAN setup prior to injecting them into a Penning trap~\cite{Bro09}. Furthermore, the BN gate is used to isolate ions with certain charge-to-mass ratios $q/m$ after charge breeding in an EBIT~\cite{Lap10}. For highly charged ions fast switching times are of particular importance to separate different ion species with very similar $q/m$. A time-of-flight spectrum of radioactive $^{76}$Rb in different charge states is shown in Fig.~\ref{fig:ToFRb76} along with residual background contamination.

\begin{figure}
   \begin{center}
   \includegraphics[width=.8\textwidth]{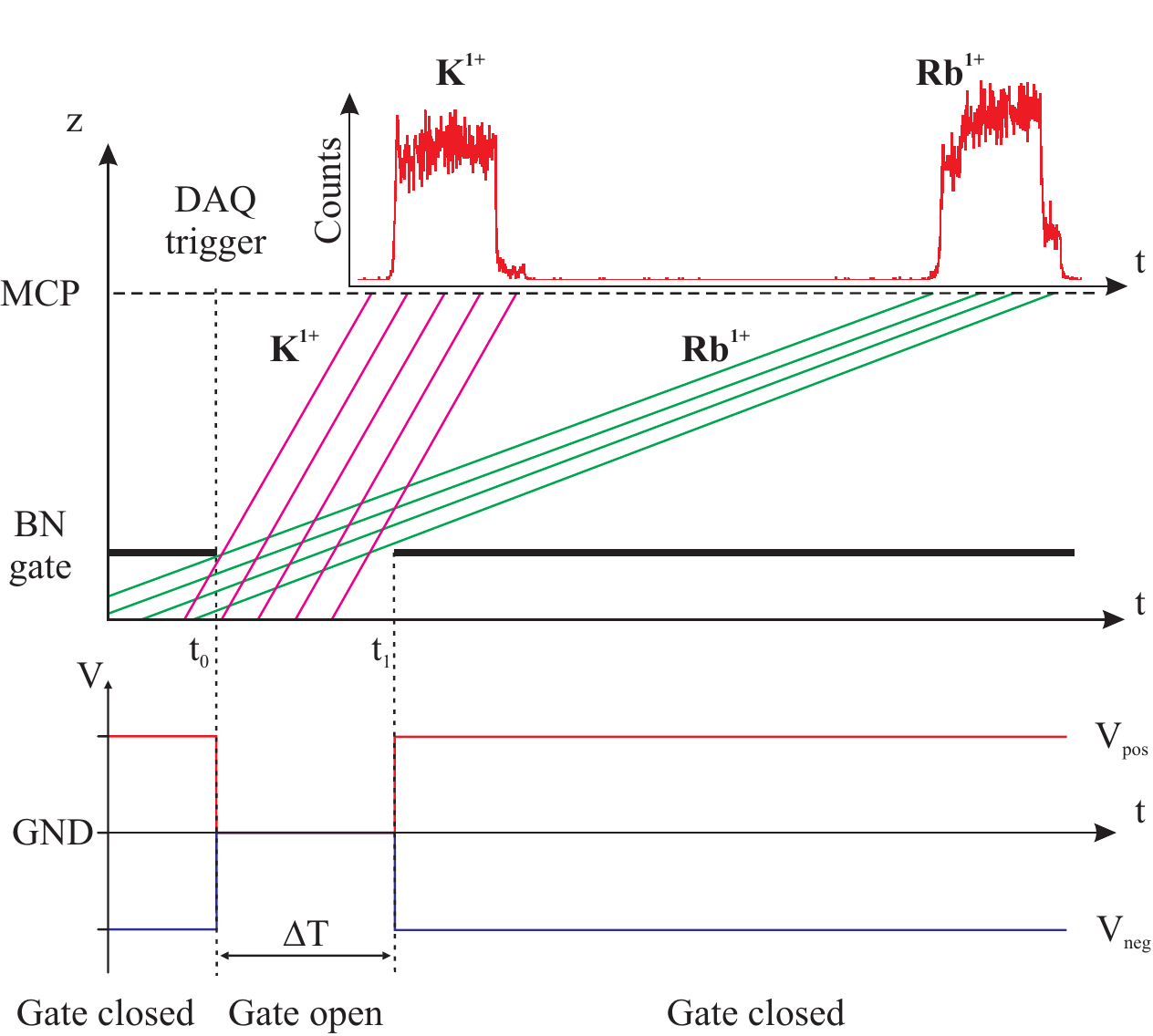}
   \end{center}
  \caption{Timing diagram used to characterize the ion gate.  The voltage on the gate and its resulting state are shown at the bottom. The top graph displays the ion's flight distance $z$ as a function of the time $t$, and the resulting time-of-flight spectrum (top). The slope corresponds to the ion's velocity. The ion source provided a continuous beam of singly-charged alkali ions ($^{39,41}$K$^+$ and $^{85,87}$Rb$^+$) with a kinetic energy of 3\,keV. Only ions reaching the MCP detector, i.e., passing the gate in its open state, are displayed in this graph.}
 \label{fig:BNG-schematic}
\end{figure}

For the characterization presented here, a continuous beam of singly charged $^{39,41}$K and $^{85,87}$Rb ions was extracted from a test ion source through a radio-frequency quadrupole cooler trap, the TITAN RFQ~\cite{Bru11a}, at a beam energy of 3\,keV. This beam was then sent towards a micro-channel plate (MCP) ion detector~\cite{Wiz79}. The BN gate was installed between the RFQ and the MCP detector at a distance of $\sim$1.35\,m from the MCP detector as illustrated in Fig.~\ref{fig:BNG-TITAN}. For this work, the gate was generally closed and only switched to transparent state (\textit{gate open}) for a short, well-defined time $\Delta$T. The voltage was switched by two solid state devices developed at TRIUMF. These switches were set up to output either the two voltages $V_{pos}$ and $V_{neg}$ to block the beam, or 0\,V to let ions pass through the gate. The rise times at the output of the switches were $\sim24$\,ns for a voltage set of $\pm160$\,V. A TTL control signal was provided to the switches by a Tektronix pulse generator (model AFG 3022B) with a rise time of 18\,ns. A synchronized TTL signal was then used to trigger the data acquisition, a multi-channel scaler (MCS Stanford Research SR 430). For the characterization of the gate, the ion's time of flight distribution was recorded varying either the applied voltages $V_{pos}$ and $V_{neg}$ or the opening time $\Delta$T. Typically, the ion's time of flight was recorded for 20,000 switching cycles at a switching rate of 500\,Hz. A schematic of the applied voltages $V_{pos}$ and $V_{neg}$ is displayed at the bottom of Fig.~\ref{fig:BNG-schematic}. During all studies, the gate assembly displayed in Fig.~\ref{fig:realgate} was installed in a four way cross CF 8'' and connected to the switch by $\sim$10\,cm long standard RG58 SHV cables. 

\begin{figure}
   \begin{center}
   \includegraphics[width=1\textwidth]{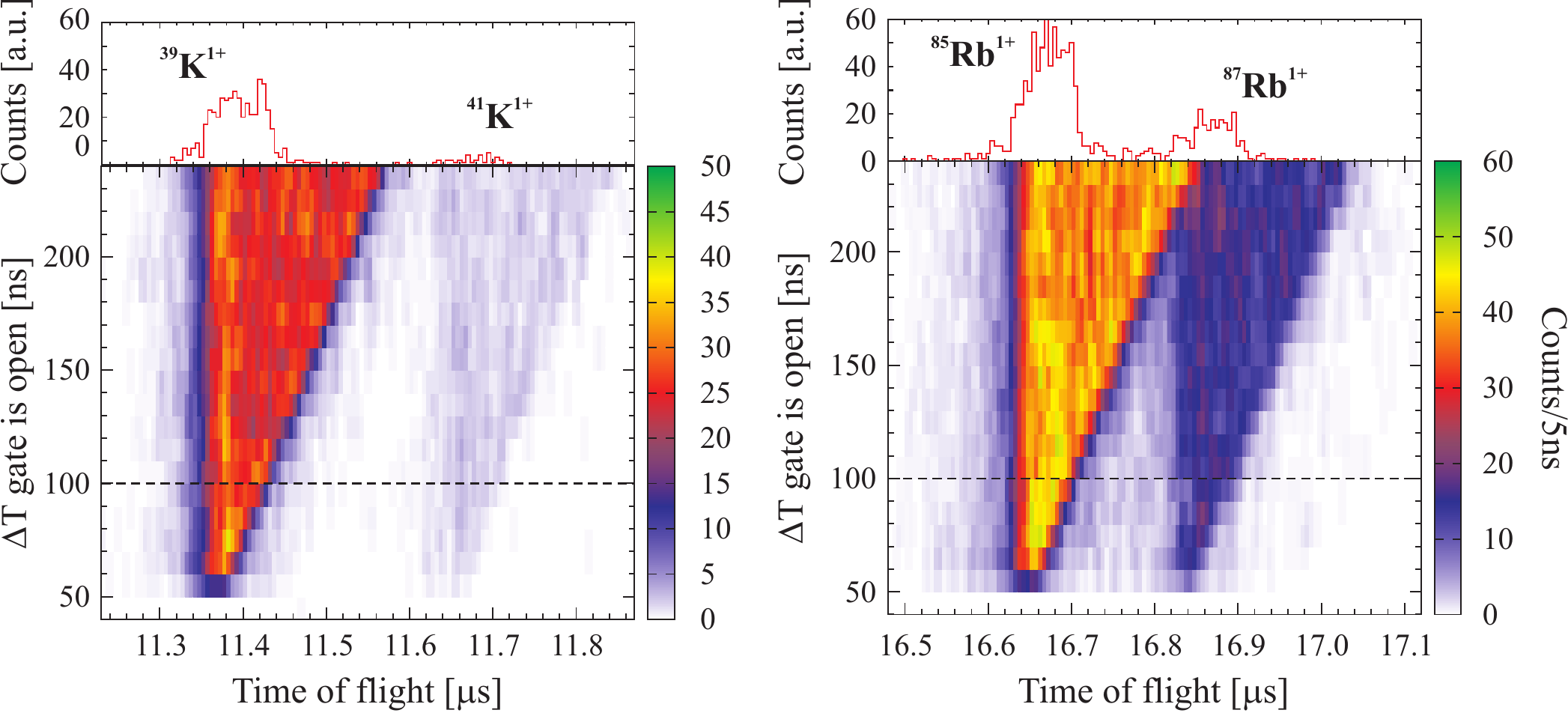}
   \end{center}
  \caption{Ion intensity as a function of $\Delta$T and the ion's flight time. K is presented on the left and Rb on the right. The time of flight spectrum for $\Delta\textnormal{T}=100$\,ns is presented at the top.}
 \label{fig:gateWidth}
\end{figure}

During these studies, grids with $51$\,\textmu m diameter ($2\,R$) and $2.2$\,mm spacing ($2\,d$) were used, resulting in a transmission of 95\%. The incoming beam with an energy of 3\,keV had an intensity of $\sim$700\,k ions/s dominantly consisting of $^{39,41}$K and $^{85,87}$Rb. Total and individual ion beam intensities were extracted from Fig.~\ref{fig:gateWidth} and the relative values for the isotopes of each element are found to agree with natural abundances (see Tab.~\ref{tab:IonIntensities}). Fig.~\ref{fig:BNG-schematic} illustrates schematically the time-of-flight measurement. Here, a steeper slope corresponds to an ion with higher velocity. At a time $t_0$ the gate is opened and ions pass through for a time period $\Delta$T. Lighter nuclides arrive earlier on the detector due to their higher speed, as shown in the figure. 

\begin{table}
\centering
 \begin{tabular}{l c c c}
  Isotope&Intensity [k ions/s]&Rel. abundance&Lit. rel. nat. abundance\\
  \hline
  $^{39}$K&261.3(5)&89(19)\%&93.2690(44)\%\\
  $^{41}$K&30.0(2)&10(2)\%&6.7310(44)\%\\
  $^{85}$Rb&422.5(7)&73(8)\%&72.165(20)\%\\
  $^{87}$Rb&153.8(4)&27(3)\%&27.835(20)\%\\
 \end{tabular}
\caption{Isotope intensities of the incoming continuous beam and the resulting relative abundances of $^{39,41}$K and $^{85,87}$Rb for comparison~\cite{ToI}.}\label{tab:IonIntensities}
\end{table}
To investigate the minimal time the gate can be opened, a symmetric voltage of $\pm180$\,V was applied to the gate and switched to ground potential for varying times $\Delta$T from 40\,ns to 240\,ns in steps of 10\,ns. The result of this measurement is presented in Fig.~\ref{fig:gateWidth} where the ion intensity at the MCP detector is displayed as a function of $\Delta$T and the ion's time of flight. For $\Delta\textnormal{T}< 50$\,ns no ions reached the MCP detector, while 
for $\Delta\textnormal{T}\lesssim 200$\,ns the incoming $^{85,87}$Rb isotopes were separated in time of flight. The time of flight spectrum for $\Delta\textnormal{T}=100$\,ns is presented at the top of Fig.~\ref{fig:gateWidth}. This value for the time has been used in the following studies, unless stated otherwise.

\begin{figure}
   \begin{center}
   \includegraphics[width=.7\textwidth]{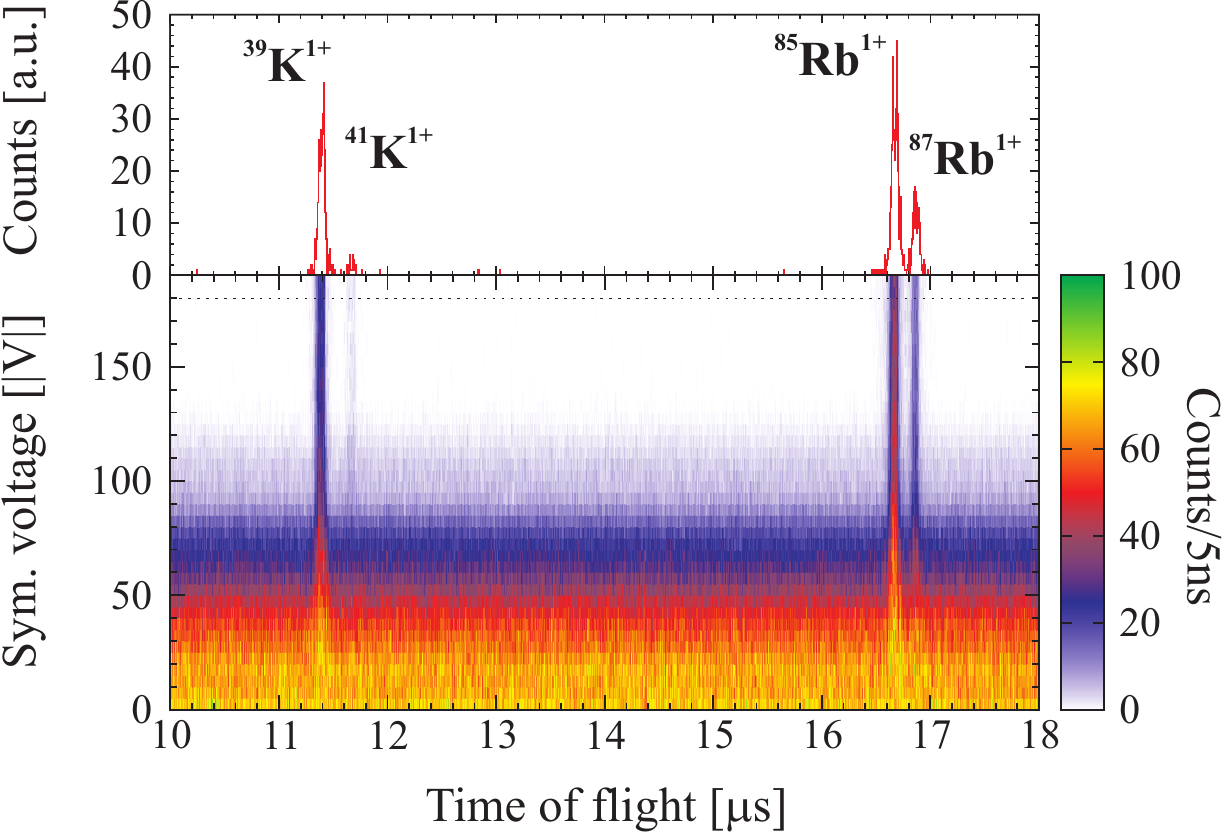}
   \end{center}
  \caption{Symmetric voltage scan with the gate open for $\Delta$T=100\,ns (bottom). The ion time-of-flight distribution is shown for $V_{pos}=180$\,V and $V_{neg}=-180$\,V (top). This configuration was used in Fig.~\ref{fig:gateWidth}.}
 \label{fig:ScanSym}
\end{figure}

In order to determine the voltage required to deflect the ions sufficiently so that they are unable to reach the MCP detector, the voltage at the wires was increased symmetrically in steps of $\pm5$\,V. The resulting scan is shown in Fig.~\ref{fig:ScanSym}. For voltages higher than $\pm120$\,V the gate is fully closed, with a negligible background (on/off ratio of $3\cdot10^{-5}$).

\begin{figure}
   \begin{center}
   \includegraphics[width=1\textwidth]{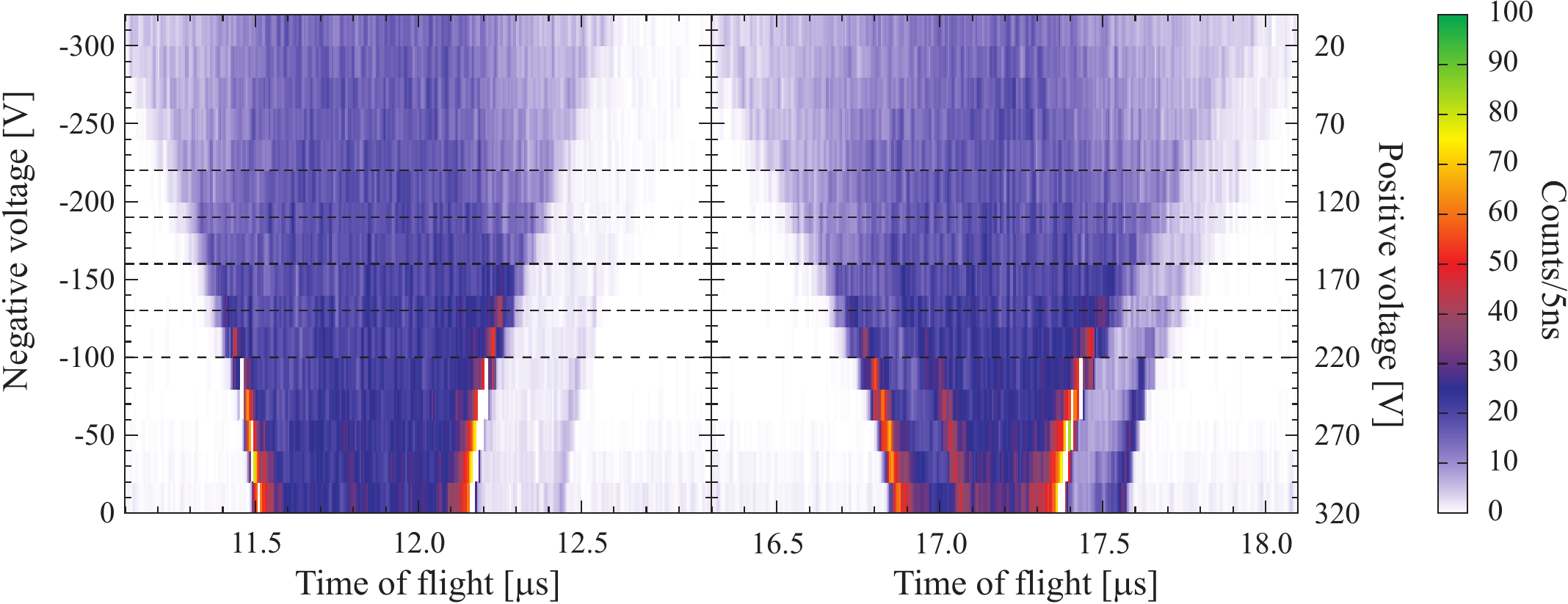}
   \end{center}
  \caption{Intensity plot of K (left) and Rb (right) for the asymmetric voltage scan. During the scan the total voltage difference between neighboring wires was kept constant at 320\,V. The gate was open for 1000\,ns. The ions' time-of-flight distribution for the voltage sets indicated by the dashed lines are presented in Fig.~\ref{fig:Tof-high-res}.}
 \label{fig:ScanAsym0-320}
\end{figure}
Additional studies were performed to investigate the concept of switching only one set of wires with twice the amplitude. This configuration could save cost of a second switch and power supply. An asymmetric voltage was applied to the wires in such a way that the potential difference between neighboring wires stayed constant at $\Delta\textnormal{V}=320$\,V. The scan started with $V_{pos}=320$\,V and $V_{neg}=0$\,V applied to the wires decrementing the voltage in steps of 20\,V, and ending at $V_{pos}=0$\,V and $V_{neg}=-320$\,V. During this scan the gate was open for $\Delta T=1000$\,ns. The resulting intensity distribution is presented in Fig.~\ref{fig:ScanAsym0-320} where the vertical axis on the right (left) displays the positive (negative) voltage applied. At $\pm160$\,V the applied voltage is symmetric. If the net voltage potential $V_{net}=V_{pos}+V_{neg}$ is negative, the ion's time of flight is smeared out while the opposite effect is observed for positive $V_{net}$. In this case, the time of flight distribution is compressed with two intense peaks at both sides of the distribution. Due to the long opening time $\Delta$T=1\textmu s, the peaks of the different isotopes overlap but can still be clearly identified.

\begin{figure}
   \begin{center}
   \includegraphics[width=1\textwidth]{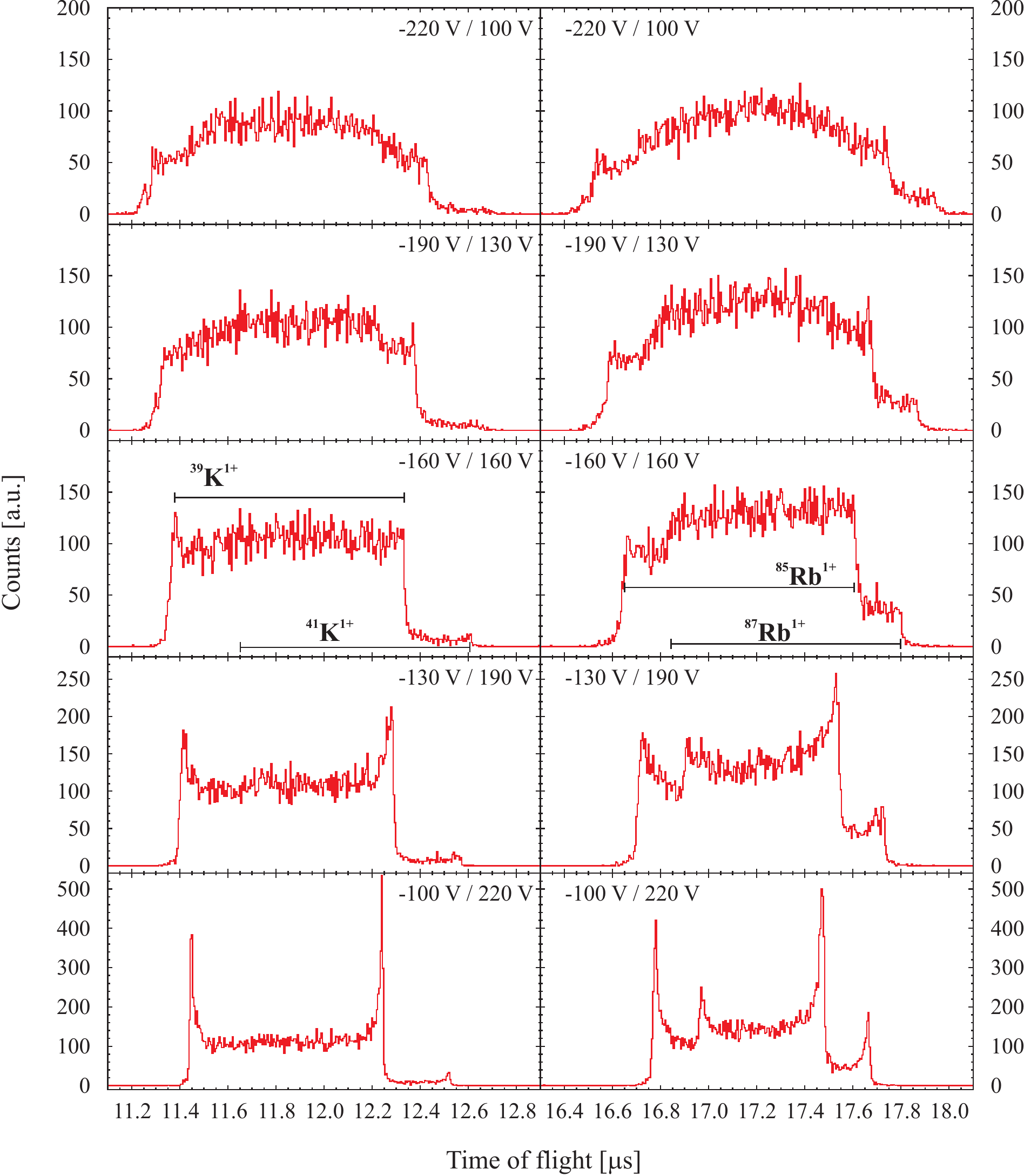}
   \end{center}
  \caption{Time-of-flight spectra for K and Rb ions for different switching voltages. The gate was open for 1000\,ns and the voltage difference between neighboring wires was kept constant at 320\,V.}
 \label{fig:Tof-high-res}
\end{figure}

\begin{figure}
   \begin{center}
   \includegraphics[width=1\textwidth]{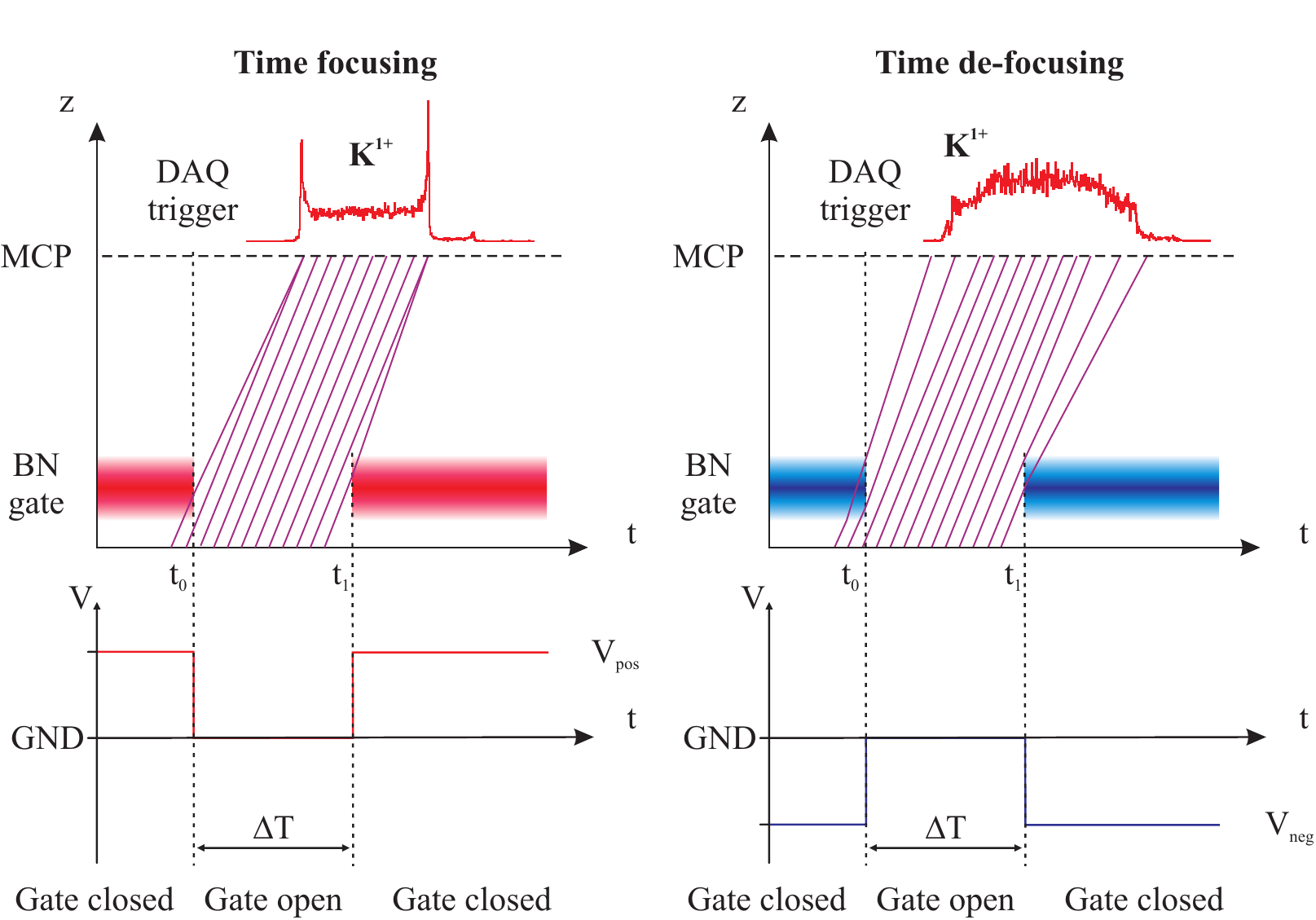}
   \end{center}
  \caption{Schematic illustrating the principle of the time focus. The timing of the drive voltages is shown at the bottom. The top plots display the ion's position along the beam axis $z$ versus time $t$. The slope represents the ion's velocity. Only K ions are displayed. The red (blue) box represents the positive (negative) net potential reaching into the space around the gate.}
 \label{fig:schematic-time-focus}
\end{figure}

In order to investigate this focusing effect, we recorded time-of-flight spectra for five voltage settings with higher statistics. Starting with the symmetric configuration of $\pm160$\,V, the voltage was incremented (or decremented) by 30\,V and 60\,V. The resulting time-of-flight distributions of K and Rb are presented in Fig.~\ref{fig:Tof-high-res}. This clearly shows the defocussing effect for $V_{net}<0$\,V and the focusing effect for $V_{net}>0$\,V. The explanation of this effect is presented in Fig.~\ref{fig:schematic-time-focus}. For the symmetric configuration, the potential is different from zero only in a very small region around the wires on the order of the wire spacing $d$. For asymmetric settings, the field originating from the applied potential $V_{net}$ acts as an advancing or retarding potential for the ions. This is illustrated by colored bars in Fig.~\ref{fig:schematic-time-focus} (red for positive and blue for negative net potential). Ions in the vicinity of this potential while the gate is switched encounter an acceleration or deceleration, depending on $V_{net}$ and their position along the beam axis when the gate is switched, as indicated by the changed slope in Fig.~\ref{fig:schematic-time-focus}. Positive ions are focused in time by $V_{net}>0$ while $V_{net}<0$ defocuses them, in both cases with the effect of broadening the energy distribution. 

\begin{figure}
   \begin{center}
   \includegraphics[width=1\textwidth]{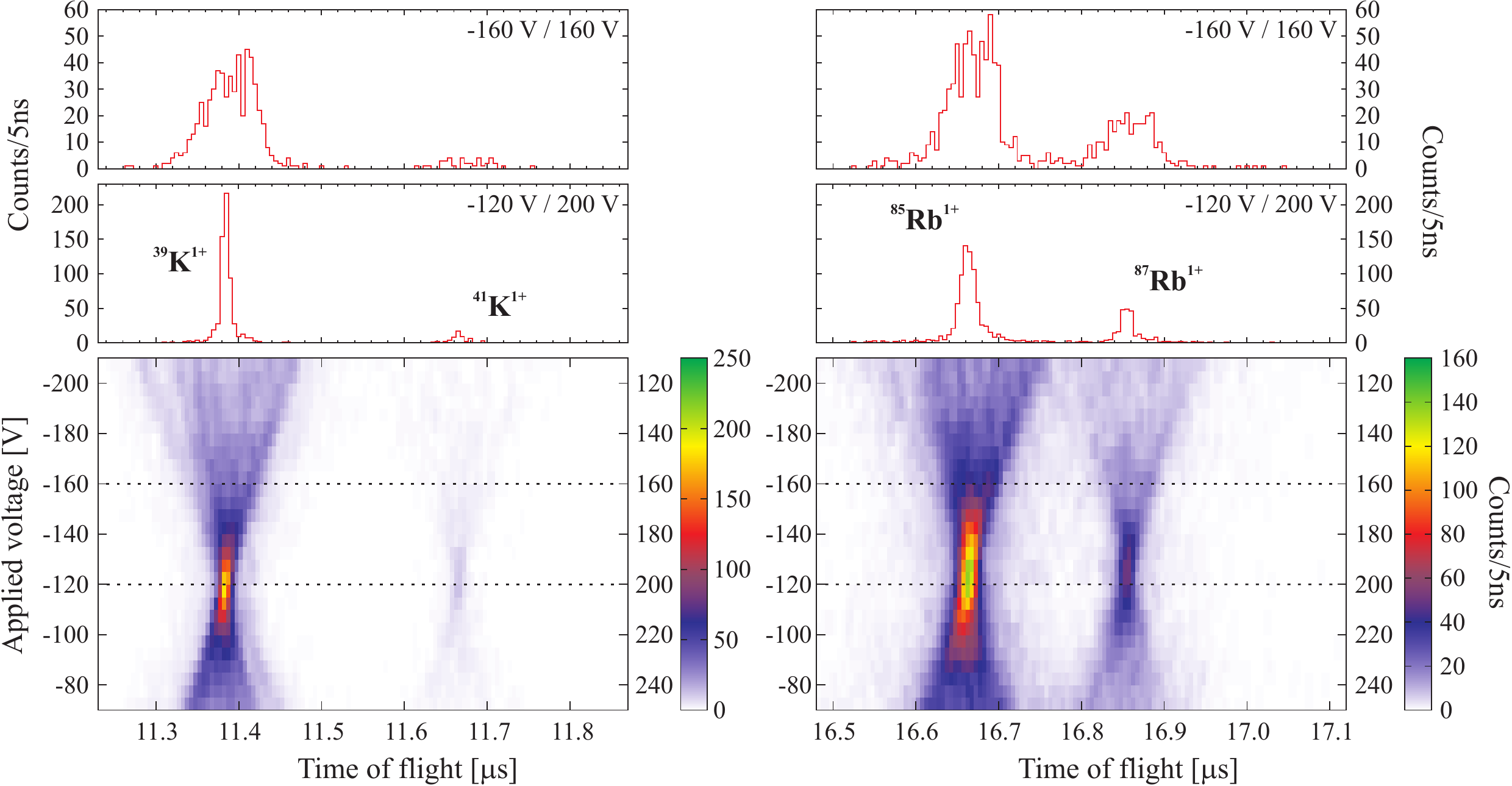}
   \end{center}
  \caption{Asymmetric scan of the voltage with a gate open for $\Delta$T=100\,ns. The dashed lines indicate the symmetric case (top) and the time-focused case (middle).}
 \label{fig:ScanAsym70-210}
\end{figure}

This focusing behavior can be used to increase the time of flight separation of the ions as shown in Fig.~\ref{fig:ScanAsym70-210}. In this scan, the gate was open for $\Delta$T=100\,ns and the voltage was scanned from $V_{neg}=-70$\,V to -210\,V and $V_{pos}=250$\,V to 110\,V in steps of 5\,V. The voltage difference was kept constant at 320\,V with the symmetric configuration being at $\pm160$\,V. At a configuration of $V_{pos}=200$\,$V_{neg}=-120$\,V the ions are focused in time maximally, therefore leading to an increase in the time-of-flight resolution. However, this has the effect of broadening the ions' energy distribution. Thus, an asymmetric voltage configuration cannot be used in the assistance of high-precision mass measurements with the TITAN Penning trap spectrometer.
\section{Conclusion}
In this publication, we present a novel design of the Bradbury Nielsen ion gate along with systematic studies of its performance that were carried out at the TITAN facility. The resulting gate has good mechanical tolerances, it is rugged and simple to assemble. In addition, it is fully UHV compatible and the wire spacing can be rapidly changed to reconfigure the device. The dimensions of the gate are scalable and so far two versions with wire grid areas of 144\,mm$^2$ and 900\,mm$^2$ have been realized. Larger areas can be realized by adapting the wire design.

Before the BN gate was installed along the TITAN beam line, the common plate of an x-y steering element was used as an ion gate to deflect the incoming beam. With this setup, the \textit{gate open} time $\Delta$T had to be greater than 300\,ns. Using the BN gate, we could improve this time by a factor of 6 allowing substantially better separation of different isotopes by their time of flight in the presented setup. This improvement is due to a lower capacitance of the BN gate ($\sim$20\,pF compared to the $\sim$70\,pF kicker) and the more favorable geometry. Even shorter \textit{gate open} times $\Delta$T might be possible with a faster electrical switch. Our measurements in the TITAN setup clearly show the applicability of this new BN gate concept for online applications with short-lived isotopes. 

Finally, we investigated the effects of asymmetrical potentials applied to the wire sets and found strong time focusing and defocusing effects. In order not to increase the energy distribution of an ion beam, symmetric potentials have to be used. However, the time focusing effect can be exploited to increase resolution on the detector
\section{Acknowledgments}
This work was supported, in part, by the US Department of Energy, the US National Science Foundation, the Natural Sciences and Engineering Research Council of Canada and the Canadian National Research Council. One of the authors (TB) acknowledges support by evangelisches Studienwerk e.V. Villigst. S.E. acknowledges support from the Vanier CGS program and M.K. was funded by the DAAD RISE program. We would like to thank M. Solyali for the construction of the gate.



\bibliographystyle{elsarticle-num}
\bibliography{BNpaper}



%

\end{document}